\documentstyle[12pt]{article}
\newcommand{\pr}{\paragraph{}}
\newcommand{\be}{\begin{equation}}
\newcommand{\ee}{\end{equation}}
\newcommand{\bea}{\begin{eqnarray}}
\newcommand{\eea}{\end{eqnarray}}
\newcommand{\nd}[1]{/\hspace{-0.6em} #1}
\newcommand{\nk}{\noindent}
\begin{document}

\begin{titlepage}

\begin{flushright}
CERN-TH.6596/92 \\
ACT-18/92 \\
CTP-TAMU-59/92
\end{flushright}
\begin{centering}
\vspace{.1in}
{\large {\bf Testing Quantum Mechanics in the Neutral Kaon System }} \\
\vspace{.4in}
{\bf John Ellis}, {\bf N.E. Mavromatos} and
{\bf D.V. Nanopoulos}$^{\dagger}$ \\

\vspace{.05in}
Theory Division, CERN, CH-1211, Geneva 23, Switzerland. \\
\vspace{.05in}
\vspace{.1in}
{\bf Abstract} \\
\vspace{.05in}
\end{centering}
{\small The neutral kaon system is
a sensitive probe of
quantum mechanics. We revive a parametrization of
non-quantum-mechanical effects that is motivated by considerations
of
the nature of space-time foam, and show how it can be constrained by
new measurements of $K_L \rightarrow 2\pi$ and $K_{L,S}$
semileptonic decays at
LEAR or a $\phi$ factory.}
\paragraph{}
\par
\vspace{0.6in}
\begin{flushleft}
CERN-TH.6596/92 \\
ACT-18/92 \\
CTP-TAMU-59/92 \\
July 1992 \\
\end{flushleft}
\vspace{0.4in}

\noindent $^{\dagger}$ {\it Permanent address}:
Center for Theoretical Physics, Dept. of Physics, \\
Texas A \& M University, College Station, TX 77843-4242, USA,
and \\
Astroparticle Physics Group,
Houston Advanced Research Center (HARC),
The Woodlands, TX 77381, USA\\

\end{titlepage}
\newpage
\section{Introduction and Summary}
\pr
    The neutral kaon system is a textbook example of a microscopic
quantum-mechanical system that exhibits a rich variety of physical
phenomena. Its resolution of the tau-theta puzzle was a manifestation
of parity violation \cite{PV}. It
is still the only place where CP violation
has been observed in the laboratory \cite{CP}, and
the suppression of $K_L \rightarrow \mu ^+\mu ^- $
decays was one of the primary motivations for charm \cite{CH}.
It
offers one of the most
sensitive available tests of the CPT invariance that is inherent to
local quantum field theory \cite{CPT}. It has provided very elegant
quantum-mechanical interference effects \cite{BS}.
Indeed, several years ago, two
of us (J.E. and D.V.N.) argued in a paper with Hagelin and Srednicki
\cite{ehns} that
the neutral kaon system was, together with long-range neutron
interferometry, one of the two most sensitive probes of a possible
breakdown of conventional quantum mechanics suggested by
investigations of local field theory in the presence of microscopic
event horizons.
\pr
It was observed some time ago \cite{hawk}
that black holes apparently
required a mixed statistical description, understood intuitively as
being due to the loss of information across the event horizon.
Hawking later suggested \cite{hawk2} that pure initial states could
evolve into mixed final states in the presence of a microscopic event
horizon. He proposed a density matrix formalism in which
$\rho_{in}$
and $\rho_{out}$ were linearly related by a $\nd{S}$-matrix
that
could not be factorized as the product of $S$- and $S^{\dagger}$-matrix
elements as expected in conventional local quantum field theory:
\be
 \rho_{out}  =  \nd{S}  \rho_{in} \qquad : \qquad  \nd{S}
 \ne S  S^{\dagger}
\label{dols}
\ee

\nk Ref. \cite{ehns} then
pointed out that in this case the normal Liouville equation
that describes the time-evolution of the quantum-mechanical density
matrix would also require modification by the addition of an extra
linear term:
\be
       \partial _t \rho    = i [ \rho ,  H  ] +  \nd{\delta H}
       \rho
\label{liouv}
\ee

\nk It
was shown explicitly that addition of the $\nd{\delta H}$
term would
allow an initially pure state to evolve into a mixed state with
positive entropy. The extra term in equation (2)
is characteristic of open quantum mechanical
systems \cite{open}.
However, we
regarded it
as necessary because of the
intrinsic impossibility of measurements within the event
horizon.
Bounds on this type of non-quantum-mechanical
behaviour were derived from the agreement of observations of neutral
kaons and of neutrons with conventional quantum mechanics. Both
systems were used \cite{ehns} to
obtain similar upper limits on the hadronic matrix
elements of $\nd{\delta H}$ of order $10^{-20}$  $GeV$.
\pr
    Although very small in a microscopic system, such effects would
be magnified in a macroscopic system with Avogadro's number of
elementary particles \cite{emhn}. They
could even engender the transition from
quantum-mechanical to classical behaviour in large systems
\cite{emhn}. Indeed,
the modification (2) has a form similar to that postulated for this
purpose in ref. \cite{ghirardi}
without any microscopic justification. Thus the
modification (2) constitutes a possible realization of the idea,
advocated more recently by Penrose \cite{penrose}, that
quantum gravity might
explain the classical behaviour of large systems. It could be
interesting to test this possibility in macroscopic
quantum-mechanical laboratories such as SQUIDs \cite{emhn}.
\pr
    Lately, the possibility of a microscopic violation of the laws of
quantum mechanics has been re-examined in the context of string
theory \cite{emn1}.
Specifically, studies of scattering and decay processs in a
spherically-symmetric string black hole background have not revealed
any loss of quantum coherence \cite{others}. We
have attributed this to the presence
in string theory of an infinite set of local symmetries that include a
$W_{1+\infty}$-algebra \cite{emn1}. This
in turn contains an infinite-dimensional
Cartan subalgebra of charges that are in involution with the
Hamiltonian, and hence conserved. Thus they provide an infinite set of
W-hair that characterizes the black hole state, preserves information,
and hence maintains quantum coherence. Thus we find no evidence for
the modifications (1,2) of conventional quantum mechanics in
scattering off
one
particular topologically non-trivial space-time background.
\pr
However, this does not mean that the $S$-matrix
of quantum field theory and conventional quantum
mechanics are sacrosanct. The symmetries that preserve
quantum coherence relate states with different
masses \cite{emn1} :
in particular, they relate the light particles
that appear in laboratory experiments to Planck
mass string states. Since realistic measurements
are conducted with a truncation of the full physical
string spectrum, they do not include all observables.
The connections between light and massive states
mean that the former should be considered as an open
system as in equation (2),
with the possibility of apparent information
loss \cite{emnprevious}. Thus it
is relevant to test the general formalism (2) also in the
context of string theory.
\pr
    Two new experimental tools to do this in the neutral kaon system
have become available since ref. \cite{ehns} was
written. One is the CP-LEAR
experiment \cite{cplear},
in which copious tagged $K^0$ decays are available, and
the other is the DA$\phi $NE $\phi$-factory now under construction
\cite{daphne}, which will
provide copious coherent $K$-${\overline K}$ pairs.
In both experiments, it
will be possible to observe CP-violating asymmetries in $K_S$ decays,
and hence new tests of CPT invariance can be
made \cite{peccei}. The purpose of
this paper is to point out that the types of measurements proposed as
tests of CPT invariance also serve as probes for violations of quantum
mechanics.
\pr
    In section 2 we remind the reader of basic features of the
modification (2) of quantum mechanics, with reference to the neutral
kaon system in which $\nd{\delta H}$ has three possible matrix elements
to be bounded by experiment \cite{ehns}. Then,
in section 3 we show explicitly how
two of them can be disentangled by measurements of CP-violating
$K_{L,S}$ semileptonic decay asymmetries and $K_L   \rightarrow   2\pi$
decays. One
of these parameters bears a phenomenological resemblance to the
CPT-violating parameter introduced in ref. \cite{peccei},
but the other appears in
a different way. Finally, in section 4 we comment on the outlook for
such probes of quantum mechanics.
\section{Formalism for the Violation of Quantum Mechanics in the Neutral
kaon System.}
\pr
This is described in the usual quantum-mechanical
framework \cite{BS}
by a
phenomenological Hamiltonian with hermitian (mass) and antihermitian
(decay) components:
\be
    H=\left(\begin{array}{c}  M-\frac{1}{2}i\Gamma \qquad
M_{12}^{*}-\frac{1}{2}i\Gamma _{12} \\
 M_{12} -\frac{1}{2}i\Gamma _{12} \qquad M-\frac{1}{2}i\Gamma
\end{array} \right)
\label{Heq}
\ee

\nk in
the ($K^0$, ${\overline K}^0$) basis. When $H$ is not hermitian, the
time-evolution of the density matrix $\rho$ is ordinarily given by
\be
  \partial _t \rho =-i (H\rho - \rho H^{\dagger} )
\label{four}
\ee

\nk and the state is pure if $Tr\rho^2$ =$( Tr\rho )^2$, which
it remains
forever it started pure. We define components of $\rho$ and $H$ by
\bea
\nonumber
\rho &\equiv & \frac{1}{2}\rho_\alpha \sigma _\alpha \\
H &\equiv & \frac{1}{2}h_\beta \sigma _\beta
\label{five}
\eea

\nk where
we use the Pauli $\sigma$-matrix basis, and the $\rho_\alpha $ are
real but the $h_\beta $ are complex.
\pr
It is convenient for our
subsequent discussion to use the CP eigenstate basis $K_{1,2}  =
\frac{1}{\sqrt{2}}( K^0  \pm {\overline K}^0 )$,
in which we can represent the ordinary
evolution (4) by $\partial _t \rho_{\alpha }=  h_{\alpha\beta}
\rho_\beta$ where
\be
 h_{\alpha\beta}
 =\left( \begin{array}{c}  - \Gamma \qquad -Re \Gamma _{12}
\qquad Im \Gamma _{12} \qquad 0 \\
 - Re \Gamma _{12} \qquad -\Gamma \qquad 0 \qquad -2Im M_{12} \\
   Im \Gamma_{12} \qquad 0 \qquad -\Gamma \qquad -2Re  M_{12} \\
  0   \qquad 2Im M_{12} \qquad 2 Re M_{12} \qquad -\Gamma
\end{array}\right)
\label{six}
\ee

\nk At large $t$, $\rho$ decays exponentially to
\be
 \rho \propto \left(\begin{array}{c}  1  \qquad \epsilon ^* \\
  \epsilon   \qquad |\epsilon |^2 \end{array} \right)
\label{seven}
\ee

\nk which corresponds to the usual pure long-lived mass eigenstate $K_L$,
with the CP impurity parameter $\epsilon$ given by
\be
  \epsilon =\frac{\frac{1}{2}i Im \Gamma _{12} - Im M_{12}}
{\frac{1}{2} \Delta \Gamma - i\Delta M }
\label{eight}
\ee

\nk where $\Delta M \equiv   M_L  -  M_S$ is positive and
$\Delta\Gamma \equiv  \Gamma_L  -  \Gamma_S$ is negative.
\pr
    We now consider the possible addition to $h_{\alpha\beta}$ (6) of a
modification of the form (2), which we parametrize as
$\nd{h}_{\alpha\beta}$. As discussed in ref. \cite{ehns}, we
assume that the
dominant violations of quantum mechanics conserve strangeness, in
which case $\nd{h}_{1\alpha}$ = 0, and therefore that
$\nd{h}_{0\alpha}$ = 0 to
conserve probability. One can show that $\nd{h}_{\alpha\beta}$ must
be a negative matrix, and hence in turn that $\nd{h}_{\alpha 1}  =
\nd{h}_{\alpha 0}$ = 0. Therefore we arrive at the general
parametrization \cite{ehns}
\be
  h_{\alpha\beta} =\left( \begin{array}{c}
 0  \qquad  0 \qquad 0 \qquad 0 \\
 0  \qquad  0 \qquad 0 \qquad 0 \\
 0  \qquad  0 \qquad -2\alpha  \qquad -2\beta \\
 0  \qquad  0 \qquad -2\beta \qquad -2\gamma \end{array}\right)
\label{nine}
\ee

\nk where the negativity of $\nd{h}_{\alpha\beta}$
further imposes $\alpha$, $\gamma$ $>$ 0 and
$\alpha$$\gamma$ $>$ $\beta ^2$. The equations of motion
for the components of $\rho$ are
\bea
\nonumber  \partial _t
\rho_{11} &=& -(\Gamma + Re \Gamma _{12} )\rho _{11}
-\gamma (\rho _{11}-\rho _{22} ) - 2Im M_{12} Re \rho _{12}
-(Im \Gamma _{12} + 2\beta ) Im \rho _{12} \\
\nonumber \partial _t \rho _{12} &=&
-(\Gamma - 2i Re M_{12} )\rho _{12} - 2i \alpha Im \rho _{12}
+ (Im M_{12} - \frac{1}{2}i Im \Gamma _{12} - i\beta )\rho _{11} \\
\nonumber &-& (Im M_{12}
 +  \frac{1}{2}i Im \Gamma _{12} - i\beta )\rho _{22} \\
\nonumber \partial _t \rho _{22}
&=& -(\Gamma - Re \Gamma _{12} )\rho _{22}
+ \gamma (\rho _{11} - \rho _{22} ) + 2 Im M_{12} Re \rho _{12} \\
&-& (Im \Gamma _{12} - 2\beta )Im \rho _{12}
\label{ten}
\eea

\nk and
it is clearly possible in principle to determine independently the
three parameters $\alpha$, $\beta$, $\gamma$ by measurements of the
evolution of the density matrix over all times.
\section{Constraining the Parameters that violate Quantum Mechanics}
\pr
Since the time-evolution (10) is described by a $4 \times 4$
linear matrix
equation, the general solution can be written in the form
\be
 \rho_\alpha (t)
 = \sum_{j=1}^4  c_{\alpha j} exp(\lambda_j t)
\label{eleven}
\ee

\nk where the coefficients $c_{\alpha j}$ depend on the initial
conditions, e.g., tagged $K_0$ or ${\overline K}_0$ beam. However, it is
clear that the large-time behaviour is dominated by the eigenvector
whose eigenvalue $\lambda _j$ has the least negative real part,
corresponding to the conventional $K_L$ component. On the other
hand, the eigenvector whose eigenvalue has the most negative real part,
corresponding to the conventional $K_S$ component, can only be probed
at short times. Interference effects at intermediate times can in
principle probe the other two eigenvectors. The feasibility
of this possibility depends crucially on the nature of the experiment,
and we will not discuss such interference
measurements here. Nor we will discuss measurements
where the correlations between $K_0$ and ${\overline K}_0 $
particles emanating from $\phi $ decay play a crucial way.
We will
concentrate on the information that
can be obtained from measurements of individual kaons
at large and small times.
\pr
    It is easy to check that, for large $t$, $\rho$ decays
exponentially to \cite{ehns}
\be
\rho \propto \left( \begin{array}{c} 1 \qquad  \qquad
\frac{-\frac{1}{2}i  (Im \Gamma _{12} + 2\beta )- Im M_{12} }
{\frac{1}{2} \Delta \Gamma + i \Delta M } \\
\frac{\frac{1}{2}i (Im \Gamma _{12} + 2\beta )- Im M_{12} }
{\frac{1}{2}\Delta \Gamma  - i \Delta M} \qquad \qquad
|\epsilon |^2 + \frac{\gamma}{\Delta \Gamma } -
\frac{4\beta Im M_{12} (\Delta M / \Delta \Gamma ) + \beta ^2 }
{\frac{1}{4} \Delta \Gamma ^2 + \Delta M^2 } \end{array} \right)
\label{twelve}
\ee

\nk where the $CP$ impurity parameter $\epsilon$ is given as usual by
equation (8). The density matrix (12) describes a mixed state with
$Tr\rho^2  < 1 $ when $\rho$ is normalized so that $Tr\rho  = 1$. It
corresponds to a {\it mixture} of a conventional $K_L$ beam with a
low-intensity $K_S$ beam. Conversely, if we look for a solution of the
time-evolution equation (10)
with $\rho_{11}  <<  \rho_{12}  << \rho_{22}$,
corresponding to what would conventionally be a $K_S$ beam, we again
find a mixed state:
\be
 \rho \propto \left( \begin{array}{c} |\epsilon |^2 +
\frac{\gamma }{|\Delta \Gamma |} -
\frac{4\beta Im M_{12} (\Delta M/\Delta \Gamma )+ \beta ^2 }
{ \frac{1}{4} \Delta \Gamma ^2 + \Delta M^2 } \qquad
\epsilon - \frac{i\beta}{\frac{\Delta \Gamma} {2} - i \Delta M} \\
\epsilon ^* + \frac{i\beta} {\frac{\Delta \Gamma}{2} +
i \Delta M} \qquad 1 \end{array} \right)
\label{thirteen}
\ee

\nk We note that the signs of the terms in (13) that are linear in
$\beta$, relative to those of $Im M_{12}$ and $Im\Gamma_{12}$, are
reversed with respect to the corresponding terms in the ``$K_L$''
density matrix (12).
\pr
    The experimental value of an observable $O$ is given in this
formalism by
\be
  \langle O \rangle = Tr  (O \rho)
\label{fourteen}
\ee

\nk as for a
conventional mixed quantum-mechanical state. The $K  \rightarrow   2\pi$
observable is represented in our $K _{1,2} $ basis by
\be
 O_{2\pi} =\left( \begin{array}{c} 0 \qquad 0 \\
0 \qquad 1 \end{array} \right)
\label{fifteen}
\ee

\nk Therefore the rate
of $K  \rightarrow   2\pi$ decays is given in the long
lifetime ``$K_L$'' limit by
\be
|\epsilon |^2 + \frac{\gamma} {\Delta \Gamma }
-\frac{4\beta Im M_{12} (\Delta M/\Delta \Gamma )+ \beta ^2 }
{\frac{1}{4} \Delta \Gamma ^2 + \Delta M^2 }
\label{sixteen}
\ee

\nk and is hence not a direct measurement of the CP-violating parameter
$\epsilon$. Previously \cite{ehns}, we
discarded the $\beta$-dependent terms
in (16), assuming that $\beta$ was similar in magnitude to $\gamma$. Here
we will be more general, keeping the term in (16) that is linear in
$\beta$.
\pr
    Other observables that are useful in constraining theories of CP
violation and, in our case, looking for a deviation from quantum
mechanics, are semileptonic $K^0/{\overline K}^0$ decays.
The $K  \rightarrow  \pi ^-l^+\nu$
observable is
\be
  O_{\pi ^-l^+ \nu} = \left(\begin{array}{c} 1 \qquad 1 \\
1\qquad 1 \end{array}\right)
\label{seventeen}
\ee

\nk whilst the $K \rightarrow  \pi^+l^-{\overline \nu}  $ observable is
\be
  O_{\pi ^+l^-{\overline \nu}}  =\left( \begin{array}{c}

1 \qquad -1 \\
 -1\qquad 1 \end{array} \right)
\label{eighteen}
\ee

\nk Hence the $CP$-violating observable $\delta$ is given by
\be
\delta \equiv \frac{\Gamma (\pi^-l^+\nu) - \Gamma (\pi^+ l^-
{\overline \nu }) }{\Gamma (\pi ^- l^+ \nu ) +
\Gamma (\pi ^+ l^- {\overline \nu})}
\label{nineteen}
\ee

\nk in both the ``$K_L$'' and ``$K_S$''
limits (12) and (13). In the usual
quantum-mechanical formalism, we would simply find
\be
  \delta \simeq 2 Re \epsilon
\label{twenty}
\ee

\nk where the
phase $\phi_{\epsilon}$ of $\epsilon$
is determined with high
precision, from measurements of $\Delta\Gamma$ and $\Delta M$. However,
using equations (12, 13, 19) we find that
\be
\delta_{L,S} \simeq 2Re [\epsilon (1-\frac{i\beta}{Im M_{12}})],
2Re [ \epsilon (1 + \frac{i\beta}{Im M_{12}})]
\label{twentyone}
\ee

\nk So far, there has not been any high-statistics measurement of
$\delta_S$. Checks of the standard phenomenology of CP violation have
been made by combining measurements of
$\delta_L$ and $K  \rightarrow   2\pi$
decays in the long lifetime limit. In our case, comparing (16) and
(21), we see that
\be
 (\frac{\delta ^2}{4} - R_{2\pi} cos^2 \phi _{\epsilon}) =
-\frac{\gamma}{|\Delta \Gamma |} cos^2 \phi _{\epsilon} -
\frac{\beta}{|\Delta \Gamma |}8|\epsilon |cos^2 \phi _{\epsilon}
sin\phi _{\epsilon}
\label{twentytwo}
\ee

\nk where $R_{2\pi} \equiv $BR$(K \rightarrow 2\pi )$.
Thus measurements of these two quantities cannot determine both
$\beta$ and $\gamma$, and the basic geometry of the problem is shown in
the figure. Any discrepancy between the measurements of
``$\epsilon$'' from $K  \rightarrow   2\pi$
decays and of ``$Re$ $\epsilon$''
from semileptonic $K$ decays could be taken as evidence
against conventional quantum mechanics, but its origin would
be ambiguous. In fact, putting in
the latest experimental values \cite{PDG}
$\sqrt{R_{2\pi}} \simeq |\epsilon | = (2.265 \pm 0.023) \times 10^{-3}$,
$\phi _{\epsilon}=43.73 \pm 0.15^{o}$,
and $\delta = (3.27 \pm 0.12 ) \times 10^{-3}$,
we find,
\be
(-0.006 \pm 0.204 ) \times 10^{-6} =
-0.522 \frac{\gamma}{\Delta \Gamma} -
(6.54 \times 10^{-3} )\frac{\beta}{\Delta \Gamma}
\label{twentythree}
\ee

\nk and hence there is currently no evidence for a violation of quantum
mechanics. If we ignore the contribution of $\beta$
in (23), and use
the
experimental value $\Delta \Gamma = 737 \times 10^{-17}
 GeV $ \cite{PDG},
we find
that
\be
        \gamma =
(0.01 \pm 0.39 ) \times 10^{-6}|\Delta \Gamma | \simeq
(0.1  \pm 3  ) \times 10^{-22} GeV
\label{twentyfour}
\ee

\nk whereas we find
\be
        \beta=
(0.01 \pm 0.31) \times 10^{-4}|\Delta \Gamma | \simeq
(1    \pm 23   ) \times 10^{-20} GeV
\label{twentyfive}
\ee

\nk if we ignore the contribution of $\gamma$ in (23).
\pr
    It is possible to disentangle the parameters $\beta$ and $\gamma$
by also measuring the semileptonic asymmetry in ``$K_S$'' decays. The
geometrical r\^ ole of this measurement is also shown in the figure.
Taking the difference between the two asymmetries in (21), we find that
\be
\delta _L  - \delta _S  = \frac{8\beta} {|\Delta \Gamma |}
\frac{sin\phi _{\epsilon}}{\sqrt{1 + \sin^2 \phi _{\epsilon}}}
= \frac{8\beta}
{|\Delta \Gamma |} sin\phi _{\epsilon}
cos\phi _{\epsilon}
\label{twentysix}
\ee

\nk Therefore a measurement of the difference between the semileptonic
decay asymmetries in the long- and short-lifetime limits is directly
sensitive to the parameter $\beta$ that violates quantum mechanics.
This measurement has also been mentioned previously as a way to look
for a violation of CPT invariance \cite{peccei}. This
is hardly a coincidence, since
it is known that a breakdown of quantum mechanics leads in general to
a weakened form of the CPT theorem of conventional local field
theory \cite{AG}. On the other hand, we have seen that another
quantum-mechanics-violating parameter $\gamma$ does not have
the same
CPT-violating signature.
\pr
    As mentioned above, it is in principle possible to determine
also the parameter $\alpha$ by measurements in the intermediate
time
region where other eigenvectors of the time-evolution matrix
equation (11) play a r\^ole. Correlation measurements
may also be interesting.
However, we will not discuss these
possibilities here.
\section{Outlook}
\pr
    We have shown in this paper that the neutral kaon system is a
uniquely precise and sensitive microscopic probe for possible
violations of quantum mechanics. It is possible to set up a
theoretically-motivated and well-defined parametrization of
non-quantum-mechanical terms in the time-evolution equation for the
density matrix of the neutral kaon
system \cite{ehns}. High-precision experiments
already constrain these parameters (25), and have the exciting prospect
of further constraining them in the future (26). The interpretation of
the bound (25) would benefit from a theoretical estimate of the
likely magnitude of non-Hamiltonian matrix elements that violate
quantum mechanics. {\it A priori}, one might expect any such matrix
elements in individual
hadronic states to be suppressed by some power of
$m_{proton}/M_{Planck}$, although we
are not yet in a position to calculate them.
As we have mentioned earlier, quantum
coherence is maintained in string theory by virtue of
symmetries linking light particles
to massive states \cite{emn1}, and
such apparently non-quantum-mechanical terms can arise
when unmeasured observables associated with massive string
states are summed over \cite{emnprevious}.
Therefore we consider it very important to take a phenomenological
attitude, and analyze this possibility from a strictly experimental
point of view. The present CP-LEAR and future DA$\phi $NE
experiments are
well-placed to contribute to this programme, since they have the
possibility to measure an asymmetry in semileptonic decays in the
short life-time limit, as well as examine interference effects that can
in principle unravel all the non-quantum-mechanical parameters.
\pr
\noindent {\Large{\bf Acknowledgements}}
\pr
Our interest in this topic was revived by a talk given
by L. Maiani at the First CERN-Torino Meeting on Current
Trends in Particle and Condensed Matter Physics.
J.E. and N.E.M. thank L. Maiani for encouraging discussions
and L. Alvarez-Gaum\'e and S. Fubini for providing
the stimulating environment of this meeting.
The work of D.V.N. is partially supported by DOE grant
DE-FG05-91-ER-40633 and by a grant from Conoco Inc.

\newpage
\pr
\noindent {\Large {\bf Figure Caption}}
\pr
The geometry of the tests of quantum mechanics
proposed in this paper. The rate $R_{2\pi}$
of $K_L \rightarrow 2\pi $ decays is not just
given by the
magnitude $|\epsilon |$ of the $CP$-violating mass mixing
parameter [see equation (16)] and the $CP$-violating
$K_{S,L}$ leptonic decay asymmetries $\delta _{S,L}$ are
not just $2$Re $\epsilon $ [see equation (21)].

\end{document}